# A study of simulating Raman spectra for alkanes with a machine learning-based polarizability model

*Mandi Fang,[a,b] Shi Tang,[a] Zheyong Fan,[c] Yao Shi,[b] Nan Xu,[a,b,*] Yi He,[a,b,d,*]*

a Institute of Zhejiang University-Quzhou, Quzhou 324000, China

b College of Chemical and Biological Engineering, Zhejiang University, Hangzhou 310027, China

c College of Physical Science and Technology, Bohai University, Jinzhou 121013, China

d Department of Chemical Engineering, University of Washington, Seattle, WA 98195, USA

**Abstract**

Polarizability is closely related to many fundamental characteristics of molecular systems and plays an indispensable role in simulating the Raman spectra. However, the calculations of polarizability for large systems still suffers from the limitations of processing ability of the quantum mechanical (QM) methods. This work assessed and compared the accuracy of the bond polarizability model (BPM) and a ML-based atomic polarizability model (AlphaML) in predicting polarizability of alkanes and then also investigated the ability of simulating Raman spectra. We found that the AlphaML has appreciable advantages over the BPM in learning the polarizability in the training data set and predicting polarizability of molecules that configurational differently from training structures. In addition, the BPM has inherent disadvantages in predicting polarizability anisotropy due to many factors including large uncertainties of estimating bond anisotropy, omitting of off-diagonal parameters in the construction of the model. As a result, the BPM has larger errors than the AlphaML in the simulation of anisotropic Raman scattering. Finally, we demonstrated that both the BPM and AlphaML suffer from transference to alkanes larger than those used in the training data sets, but the problem for the AlphaML can be circumvented by exploring more proper training structures.

**1. Introduction**

Polarizability describes the tendency of an atom or molecule to adjust its electron cloud in response to an external electric field,[1,2] and therefore plays a vital role in understanding nonlinear optics and the Raman scattering[3,4]. It is also an indispensable ingredient in the development of next-generation polarizable force fields.[5,6]

Technically, polarizability can be calculated using the quantum mechanical (QM) methods such as the density functional theory (DFT).[7,8] However, the computational cost of DFT calculations is proportional to the cube of system size. The small-system-size limitations hinder the applications of the DFT method in calculating polarizabilities of systems with more than hundreds of atoms.[9–11] It is possible to circumvent the problem by developing parametric polarizability models, which have a linear scaling of the computational effort with system size.[12]

The bond polarizability model (BPM) is a frequently-used parametric polarizability model, in which the molecular polarizability is treated as the sum of polarizabilities of all chemical bonds in the systems.[11,13,14] Bougeard and coworkers[13] parameterized a BPM to predict polarizabilities for alkanes using molecular polarizabilities calculated by the DFT method as training data. The results demonstrate that the trained BPM learn polarizabilities in the data set well and enables transfers to molecules larger than those in the training data sets. Based on this line of thought, Milner and coworkers[14] retrained a BPM for alkanes, and transferred the model to polyethylene. By combining the BPM with molecular dynamics (MD) simulations, the authors simulated Raman spectra of polyethylene and successfully described the difference between Raman peaks of polyethylene at crystalline and molten phases.

In spite of successfully describing the difference in Raman peaks for polyethylene, the BPMs still have large errors when predicting the polarizabilities of alkanes,[11,13,14] which seriously questions the quality of the simulated Raman spectra. The preset functions to map from bonds' attributes (lengths, orientations etc.) to their polarizabilities may limit the ability of the BPM to describe polarizability under complex chemical environments. Machine learning-based polarizability models are considered as an alternative parametric model for predicting molecular polarizability with high accuracy. Typical packages for this purpose include the AlphaML[15], embedded atom neural network (EANN)[16] and deep neural network (DNN)[17]. In these models, chemical environment of each atom within a cutoff is encoded into mathematical descriptors, and then the descriptors are inputted into a ML model to give a polarizability tensor. Due to the specially designed descriptors and the powerful fitting capability of ML, these models are able to retain the accuracy of the underlying QM methods but can predict molecular polarizability several orders of magnitude faster.[12,15,18]

Alkanes are a class of compounds which are extensively studied in the development of the BPMs.[13,19–21] Two reasons may contribute to the popularity of alkanes. Firstly, alkanes have a large quantity of spectroscopic data available which serve as experimental references. Secondly, alkanes and the corresponding polymers (polyolefin) are well suitable to the study on the transferability of the BPMs.[11,22] Although the machine learning-based polarizability models have achieved great success in predicting polarizabilities of waters and organic compounds[15,16,18,23], few researches focus on the prediction of polarizability, transferability of models and simulation of Raman spectra for alkanes.

The aim of the present work is to assess a ML-based polarizability model in terms of the accuracy of predicting molecular polarizability and simulating Raman spectra for alkanes. We also trained a zero-order BPM for comparisons. Polarizabilities of alkanes containing no more than 5 carbons were used as the training data set and testing data set. The performance of the ML-based polarizability model on predicting polarizabilities were evaluated, and then compared with the results produced by the BPM. In addition, the Raman spectra of ethane and 2-methylbutane were simulated by combining the polarizability models with molecular dynamics (MD) simulations, and the spectra were then compared with those simulated by the combination of the DFT method and MD simulations. We further investigated the transferability of the AlphaML model and the BPM by using larger alkanes containing no more than 11 carbons as a testing data set.

## 2. Simulation Details

### 2.1 DFT calculations

Geometries of alkanes were optimized using the B3LYP exchange-correlation functional[24] and the 6-31G(d) basis set.[25] Dispersion corrections were considered with the DFT-D3 method with Becke-Johnson damping (D3BJ).[26,27] Based on the optimized structures, molecular polarizabilities were calculated by solving the CPHF equations using the B3LYP functional and the 6-311++G(2d,2p) basis set.[28] All DFT calculations were carried out using the ORCA-v5.0 package[29,30] with the RIJCOSX numerical integration.[31,32]

**2.2 Preparations of data sets**

The initial training data set consists of methane, ethane, propane, n-butane, n-pentane and 2-methylbutane. For n-butane and n-pentane, both the trans and gauche conformations were included, therefore, the initial training data set contains 8 structures in total, which is the same to the training data set for BPM in the work of Bougeard and coworkers.[13] An additional training data set containing 95 structures was constructed by stretching the individual C-C and C-H bonds in the data set by $\Delta r = 0.01 \text{ Å}$. Molecular polarizability of each structure was calculated using the method introduced in section 2.1.

A testing data set of non-equilibrium molecules was constructed to evaluate the accuracy of the polarizability models in predicting polarizability of molecules that are far away from equilibrium. Using structures of 8 alkanes at equilibrium as the initial structures, MD simulations at 300 K were performed to obtain non-equilibrium structures. The temperature was controlled using the Berendsen thermostat algorisms with the NVT ensemble.[33] Meanwhile, all C-H bonds were constrained with the SHAKE algorithm.[34] The xTB package[35] and the GFN2-xTB[36] force field were used for all MD simulations. Each simulation has a duration of 5 ps with a time step of 1 fs, and trajectories were saved every 1 ps. In this way, 40 structures were collected.

Examination of transferability was performed for alkanes containing no more than 11 carbons. These structures were firstly generated from SMILES strings in the GDB-11 database[37] using the RDKit package.[38] Then, geometry optimizations were performed with the DFT method followed by calculations of molecular polarizability.

**2.3 Principle of the ML-based polarizability model**

In the present work, we chose the AlphaML[15] as a representative of the ML-based polarizability models. The AlphaML is based on a symmetry-adapted Gaussian process regression (SA-GPR) scheme, which is designed for the predictions of tensorial properties.[23] We shall introduce the component-wise GPR firstly, which is a simplification of the SA-GPR scheme.[39] In this scheme, each individual polarizability component $\alpha_{pq}$ in the Cartesian frame ($p, q = x, y, z$) reads

$$\alpha_{pq}(\mathcal{B}) = \bar{\alpha}_{pq}^{\text{ave}} + \sum_{j=1}^{N} w_j^{pq} k(\mathcal{B}, \mathcal{A}_j) \tag{1}$$

where $N$ is the number of configurations in the training data set, $\bar{\alpha}_{pq}^{\text{ave}}$ is the average of the polarizability component $\alpha_{pq}$ over the training data set, $w_j^{pq}$ are the weights, and $k$ is a kernel function that measure the similarity between the target system $\mathcal{B}$ and a training structure $\mathcal{A}_j$. The kernel function commonly used in the GPR is based on a Gaussian similarity

$$k(\mathcal{B}, \mathcal{A}_j) = \exp\left(-\frac{|\mathbf{u}(\mathcal{B}) - \mathbf{u}(\mathcal{A}_j)|^2}{2\sigma^2}\right) \tag{2}$$

where $\sigma$ is the Gaussian width, and $\mathbf{u}(\cdots)$ is a function that maps the atomic coordinates to a high-dimensional space.

The mapping function $\mathbf{u}(\cdots)$ determines the accuracy and efficiency of the GPR model. It is often constructed using the atomic density representation, which builds a Cartesian reference frame centered on an atom and defines a three-dimensional grid around it. At each grid-point $\mathbf{r}$, the atomic density distribution is calculated in the form

$$\rho_s(\mathbf{r}) = \sum_{i \in s} \exp\left(-\frac{|\mathbf{r}-\mathbf{r}_i|^2}{2\gamma_s^2}\right) \tag{3}$$

where $s$ identifies an atom type, $\mathbf{r}_i$ is the coordinate of the central atom, and $\gamma_s$ is a smearing parameter. $\mathbf{u}(\cdots)$ is given by the set $\{\rho_s(\mathbf{r}), s = 1, \cdots, N_s\}$, where $N_s$ is the number of atomic species in the system. The polarizability tensor predicted in this manner is not invariant to rotations in Cartesian space, and therefore the atomic density that uses a Cartesian space representation requires an alignment to a reference structure.[39]

The AlphaML learns the polarizability using a combination of GPR and SA-GPR scheme[23,39]. Naturally, the polarizability $\boldsymbol{\alpha}$ is a symmetric rank-2 tensor. Before fitting, $\boldsymbol{\alpha}$ is decomposed into a scalar component $\alpha^{(0)} = (\alpha_{xx}+\alpha_{yy}+\alpha_{zz})/\sqrt{3}$ and a tensorial component $\alpha^{(2)} = \sqrt{2}\left[\alpha_{xy}, \alpha_{yz}, \alpha_{xz}, \frac{2\alpha_{zz}-\alpha_{xx}-\alpha_{yy}}{2\sqrt{3}}, \frac{\alpha_{xx}-\alpha_{yy}}{2}\right]$.[15] The former is fitted using the GPR scheme while the latter is fitted using SA-GPR scheme. Within the SA-GPR scheme, a tensorial generalization of the smooth overlap of atomic position kernel (λ-SOAP) are employed, which uses the covariant integration of the atomic density in the mapping functions. By using the λ-SOAP for the tensorial component, the AlphaML gets rid of the alignment for the atomic density. Detailed descriptions of the SA-GPR scheme and the λ-SOAP kernels can be found elsewhere.[23,39]

The training of the AlphaML model is equivalent to the selection of representative reference environments and the determination of the corresponding weights $w_j^{pq}$, which is done by minimizing a loss function defining the deviations of predicted polarizabilities from those given in the training data set.[15,39] For $\alpha^{(0)}$, 8 radial functions, 6 angular functions, an environment cutoff of 5 Å and a Gaussian width of 0.25 Å were used. For $\alpha^{(2)}$, the hyperparameters are the same as those for the scalar component except that a Gaussian width of 0.35 Å was used.

**2.4 Simulations of Raman spectra**

The isotropic and anisotropic Raman scattering are closely related to the isotropic polarizability $\bar{\alpha} = (\alpha_{xx}+\alpha_{yy}+\alpha_{zz})/3$ and the anisotropic tensor $\widetilde{\boldsymbol{\alpha}} = \boldsymbol{\alpha} - \bar{\alpha}\mathbf{I}$, respectively. Neglecting the nuclear quantum effects, the differential cross section of isotropic and anisotropic Raman scattering can be written in terms of the Fourier transform of the polarizability autocorrelation function (PACF)[4,11]

$$\left(\frac{d^2\sigma}{d\omega d\Omega}\right)_{iso} = \frac{1}{2\pi}\int_{-\infty}^{+\infty} dt e^{-i\omega t}\langle\bar{\alpha}(0)\bar{\alpha}(t)\rangle \tag{4}$$

$$\left(\frac{d^2\sigma}{d\omega d\Omega}\right)_{aniso} = \frac{1}{2\pi}\int_{-\infty}^{+\infty} dt e^{-i\omega t}\langle\text{Tr}[\widetilde{\boldsymbol{\alpha}}(0)\widetilde{\boldsymbol{\alpha}}(t)]\rangle \tag{5}$$

where $\omega$ is the Raman frequency shift, $\Omega$ is the solid angle range. Tr indicates the trace and $\langle\cdots\rangle$ indicates an ensemble average. The parallel and perpendicular spectra that are comparable to the experimental polarized Raman spectra can be obtained by[4,11,18]

$$I_{\parallel}(\omega) = \left(\frac{d^2\sigma}{d\omega d\Omega}\right)_{iso} + \frac{2}{15}\left(\frac{d^2\sigma}{d\omega d\Omega}\right)_{aniso} \tag{6}$$

$$I_{\perp}(\omega) = \frac{1}{10}\left(\frac{d^2\sigma}{d\omega d\Omega}\right)_{aniso} \tag{7}$$

It is worth mentioning that the simulated spectra only record the line shapes of Raman spectra[18], which is independent on the wavelength of the incident light.

To simulate Raman spectra for ethane and 2-methylbutane, we performed a MD simulation with the GFN2-xTB[36] force field. The simulation has a duration of 25 ps with a time step of 1 fs and trajectories were saved every 2 fs. Polarizabilities of structures sampled from the MD simulation were calculated by the DFT method, the BPM and the AlphaML model, respectively, and the corresponding PACF along the evolution of trajectories were then obtained. Finally, we simulated the parallel and perpendicular spectra according to the formulae in Equation 4~7. The Fourier transforms of PACF in Equations 4 and 5 were simplified with the use of the discrete cosine transform (DCT).[40] A combination of a sampling interval of 2 fs and a lag time of 10 ps in DCT produces an increment of 1.67 cm$^{-1}$ in frequency, which can capture the highest frequency mode of CH bending with a period of about 23 fs.[14]

## 3. Result and discussion

In the very beginning, we trained the AlphaML model and the zero-order BPM with the initial training data set containing 8 alkanes at equilibrium and 95 structures with a stretched C-C or C-H bond. We demonstrated the training process and the fitting quality of the AlphaML in the following section, while presenting the training process of the BPM in the supporting material. In the second step, comparisons were performed between the BPM and the AlphaML model in terms of the accuracy of predicting polarizabilities for molecules that are far away from equilibrium. Subsequently, Raman spectra of ethane and 2-methylbutane were simulated by integrating the AlphaML model, the BPM or the DFT method with MD simulations. The relationship between the accuracy of polarizability prediction and the quality of simulated Raman spectra were demonstrated. Finally, we investigate the transferability of the AlphaML model and the BPM to alkanes larger than those in the training data sets.

## 3.1 Training of the AlphaML model

We used the training data set containing a total of 103 structures to train the AlphaML model. **Figure 1a** shows explicitly scatter plots of the diagonal and off-diagonal components of polarizabilities calculated by the DFT method and predicted by the AlphaML model for the 8 alkanes at equilibrium. The AlphaML model yields coefficient of determination ($R^2$) near 1.00 for both the diagonal and off-diagonal components. As presented in the supporting material, the BPM yields $R^2$ of 0.997 and 0.803 for the diagonal and off-diagonal components, indicating that the AlphaML possesses a better learning capacity for the off-diagonal components of polarizabilities. Meanwhile, we also examined the ability of predicting the derivatives of polarizability associated with a bond stretch. The narrow distributions of the derivatives of polarizability around the $y = x$ line manifest that the AlphaML has enough accuracy to describe the tiny changes of polarizabilities associated with bond stretches, as shown in **Figure 1b**. The $R^2$ for both the diagonal and off-diagonal components of derivatives of polarizability are greater than 0.993, much higher than the corresponding $R^2$ produced by the BPM, as shown in **Table 1**. The result echoes the trend that AlphaML has advantage over the BPM in learning the polarizability of molecules in the condition of identical training data sets, which are consistent with what we expected due to the powerful fitting capability of ML[12].

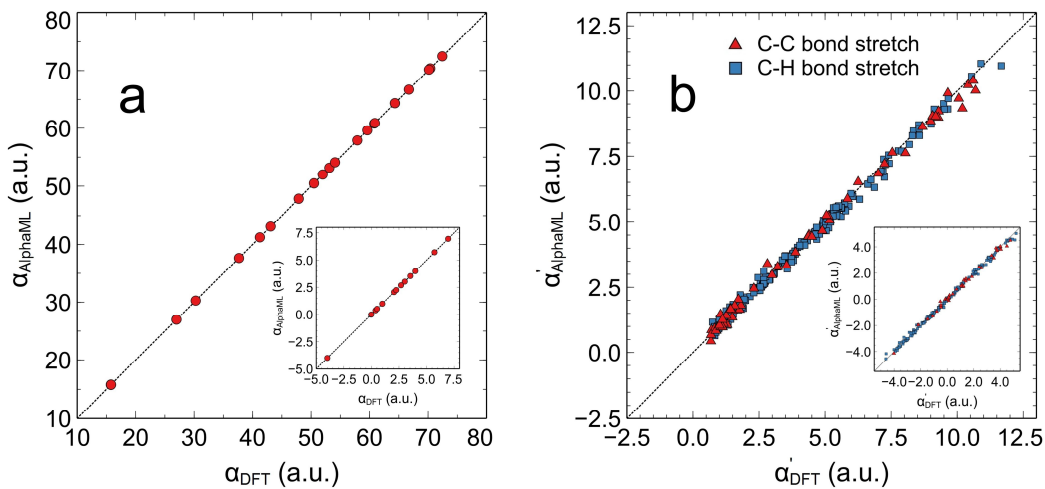

**Figure 1.** (a) Correlations between diagonal components of polarizabilities of 8 alkanes at equilibrium calculated by the DFT method and predicted by the AlphaML model. Correlations for off-diagonal components are shown in the inset. (b) Correlations between diagonal components of the derivatives of polarizability associated with a bond stretch calculated by the DFT method and predicted by the AlphaML model. Correlations for off-diagonal components are shown in the inset.

**Table 1. The $R^2$ for the diagonal and off-diagonal components of derivatives of polarizability associated with a bond stretch.**

|  | $R^2$ (diagonal/off-diagonal) | |
| --- | --- | --- |
| Model | C-C stretch | C-H stretch |
| BPM | 0.895/0.879 | 0.842/0.909 |
| AlphaML | 0.996/0.993 | 0.995/0.997 |

**3.2 Performance on predicting polarizability of non-equilibrium structures**

The AlphaML model exhibits an extraordinary ability in learning polarizabilities and derivatives of polarizability from the training data set containing structures at equilibrium and structures with small bond stretches. A benchmark testing was performed to the AlphaML model and the BPM to evaluate the accuracy in predicting polarizability of structures that are far away from equilibrium.

**Figure 2a** shows that both the BPM and the AlphaML model can predict polarizabilities of the non-equilibrium structures in agreement with the DFT results, although the BPM has larger errors in predicting the off-diagonal components than the AlphaML model. This trend is consistent with the fact that the AlphaML model can learn the off-diagonal components of polarizabilities of 8 alkanes at equilibrium better than the BPM as presented in the previous section. To rule out the effects of rotation operations on molecular polarizabilities, here, we introduced the rotation-invariant isotropic polarizability ($\bar{\alpha}$) and the rotation-invariant polarizability anisotropy[41] ($\Delta\alpha = \sqrt{\left[(\alpha_{xx}-\alpha_{yy})^2+(\alpha_{yy}-\alpha_{zz})^2+(\alpha_{xx}-\alpha_{zz})^2+6(\alpha_{xy}^2+\alpha_{xz}^2+\alpha_{yz}^2)\right]/2}$) to adequately measure the deviations of polarizability predicted by the AlphaML model and the BPM from the corresponding polarizabilities calculated by the DFT method.

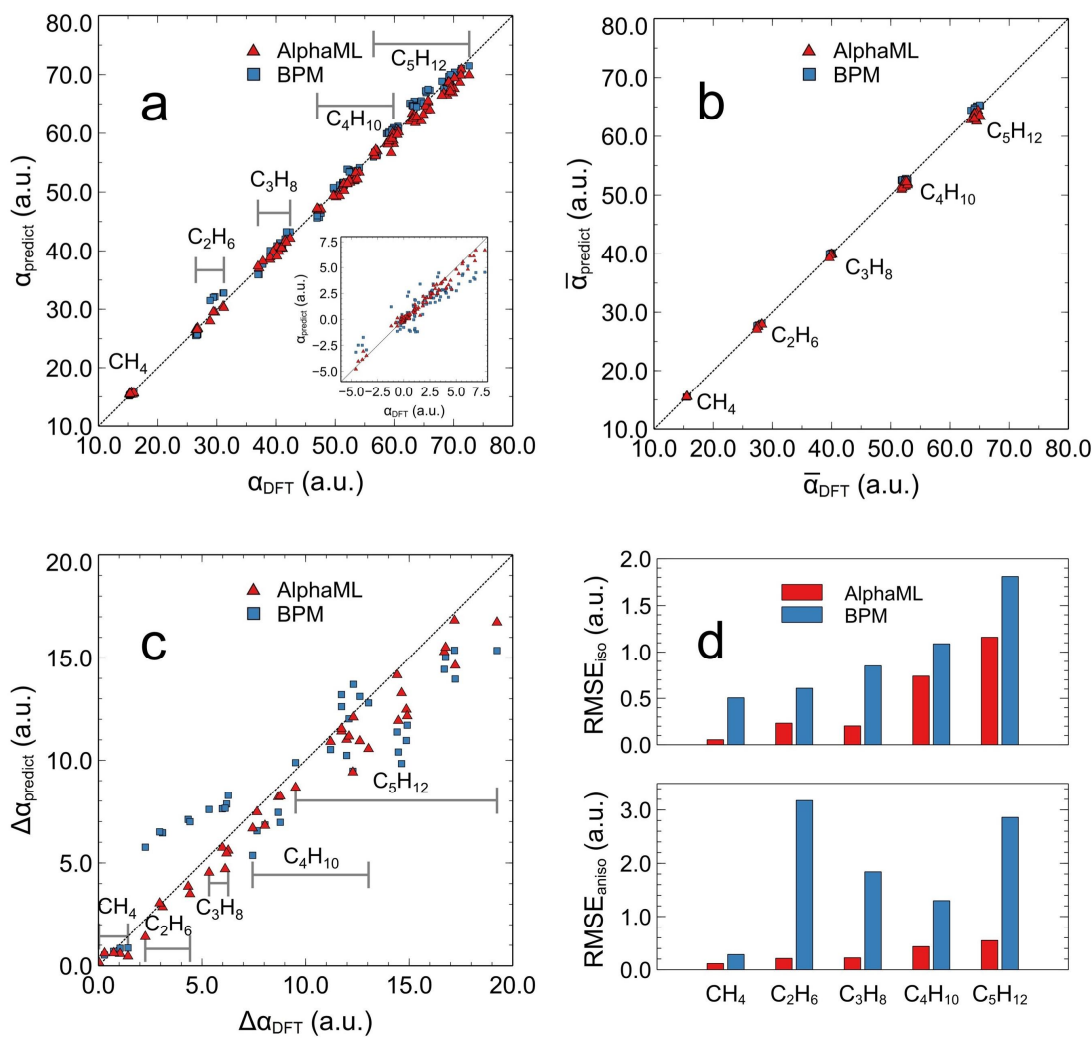

**Figure 2.** (a) Correlations between diagonal components of polarizabilities of non-equilibrium structures calculated by the DFT method and predicted by the AlphaML model and the BPM. Correlations for off-diagonal components are shown in the inset. (b) Correlations between isotropic polarizability ($\bar{\alpha}$) calculated by the DFT method and predicted by the AlphaML model and the BPM. (c) Correlations between polarizability anisotropy ($\Delta\alpha$) calculated by the DFT method and predicted by the AlphaML model and the BPM. (d) RMSEs of isotropic polarizability ($RMSE_{iso}$) and polarizability anisotropy ($RMSE_{aniso}$) produced by the AlphaML model and the BPM.

Both the BPM and the AlphaML can predict $\bar{\alpha}$ as good as the DFT method, as shown in **Figure 2b**. The results agree with the trend for the diagonal components of polarizabilities shown in **Figure 2a** well, as $\bar{\alpha}$ is the average of the three diagonal components. It is interesting that data in **Figure 2b** are grouped by their chemical formula of molecules, namely $CH_4$, $C_2H_6$, $C_3H_8$, $C_4H_{10}$ and $C_5H_{12}$. Moreover, the groups of data distribute around the $y = x$ line almost uniformly, which implies that $\bar{\alpha}$ may increase linearly with number of carbons in the molecules. This linear relationship is confirmed in **Figure S2**. In other words, the additive assumption of polarizabilities is roughly valid for the average of diagonal components. Both the BPM and the AlphaML model assume that the molecular polarizabilities are additive,[11,15] thus, the high consistency between the predicted $\bar{\alpha}$ and $\bar{\alpha}$ calculated by the DFT method is not surprising.

The AlphaML can predict $\Delta\alpha$ much better than the BPM for non-equilibrium structures. **Figure 2c** clearly shows that the points of $\Delta\alpha$ predicted by the BPM are far away from the $y = x$ line, indicating rather higher prediction errors. This trend is confirmed from high RMSEs of $\Delta\alpha$ shown in **Figure 2d**. On the contrary, $\Delta\alpha$ predicted by the AlphaML model agree well with DFT values, which is validated by the low RMSEs of $\Delta\alpha$ shown in **Figure 2d**. The points of $\Delta\alpha$ predicted by the BPM model are well correlated with those calculated by the DFT method, also having a trend of slight downward offset, as shown in **Figure 2c**. The effect of the overall offset on the quality of Raman spectra will be demonstrated in the following section. By the way, the reasons for the high RMSEs of $\Delta\alpha$ given by the BPM is deeply discussed in the supporting material.

**3.3 Performance on predicting Raman Spectra**

In the previous section, we have demonstrated the performance of the AlphaML model on predicting the isotropic polarizability and the polarizability anisotropy for non-equilibrium alkanes. The AlphaML model is proven to have better accuracy than the BPM in the condition of identical training data sets. In this section, we will evaluate the quality of simulated Raman spectra for ethane and 2-methylbutane and interpret the relationship between the accuracy of polarizability prediction and quality of simulated Raman spectra.

We calculated the time autocorrelation function of isotropic polarizability ($\text{PACF}_{iso}(t) = \langle \bar{\alpha}(0)\bar{\alpha}(t) \rangle$) and anisotropic polarizability tensor ($\text{PACF}_{aniso}(t) = \langle \text{Tr}[\tilde{\boldsymbol{\alpha}}(0)\tilde{\boldsymbol{\alpha}}(t)] \rangle$) for ethane using the AlphaML model and the BPM, and compared the time autocorrelation function with the corresponding values calculated by the DFT method. **Figure 3a** shows that the wave shape and amplitude of $\text{PACF}_{iso}$ predicted by the AlphaML model and the BPM are close to that calculated by the DFT method, which agrees with the fact that both the AlphaML model and the BPM can predict the $\bar{\alpha}$ for ethane as good as the DFT method. The $\text{PACF}_{aniso}$ predicted by the AlphaML model also resembles the $\text{PACF}_{aniso}$ calculated by the DFT method. However, the $\text{PACF}_{aniso}$ predicted by the BPM has distinctive differences from the $\text{PACF}_{aniso}$ calculated by the DFT method in terms of the wave shape and period of change, as shown in **Figure 3b**. Since the Raman scattering intensity is proportional to the Fourier transform of the PACF,[4,11] the obvious change of the wave shape and period may cause the difference in the shifts and the intensity of peaks of the simulated Raman spectra.

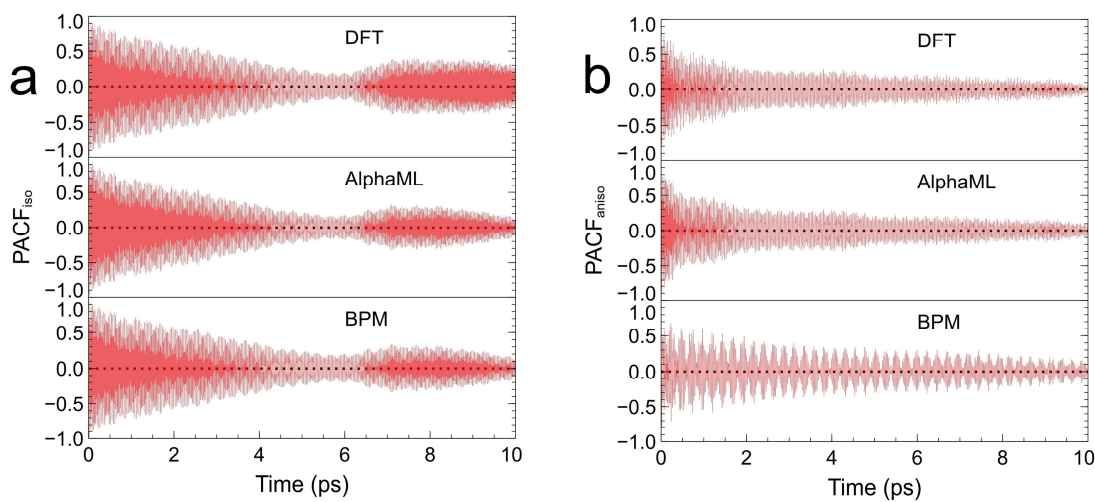

**Figure 3.** Comparison between (a) $PACF_{iso}$ and (b) $PACF_{aniso}$ for ethane calculated by the DFT method and predicted by the AlphaML model and the BPM.

The Raman spectra of ethane simulated from PACFs predicted by the AlphaML model are consistent with that simulated from PACFs calculated by the DFT method across a wide range of wavenumbers, as shown in **Figure 4**. The Raman spectra simulated from PACFs by the DFT method, the AlphaML model and the BPM can successfully reproduce an array of features reported in experiments[21,42], with coincident Raman shifts and approximately close intensities. These features include the C-C stretching peak at about 1057 cm$^{-1}$, CH$_3$ stretching peaks at about 3054 and 3067 cm$^{-1}$, as shown in **Table 2**.

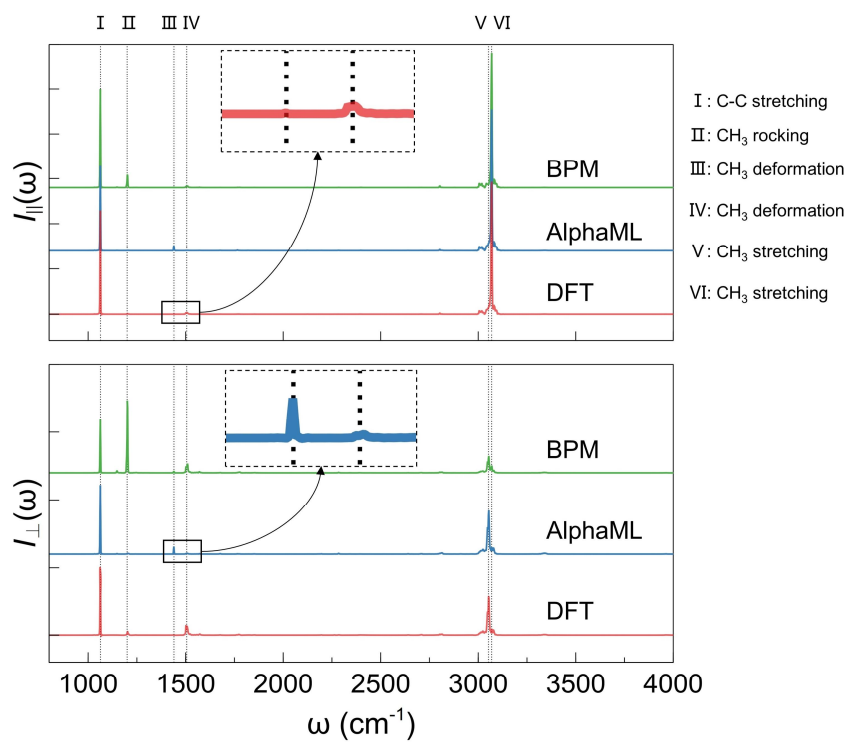

**Figure 4.** Comparisons between parallel and perpendicular Raman spectra of $C_2H_6$ simulated from PACFs by DFT, the AlphaML and the BPM. The latter two Raman spectra were shifted upward.

**Table 2. Wavenumber of the Raman Spectra of $C_2H_6$ from experiments and simulated from PACFs calculated by DFT (cm$^{-1}$)**

| skeletal mode description | experiment | DFT |
|---|---|---|
| $CH_3$ stretching | 2953.7 | 3054 |
| $CH_3$ stretching | 2968.7 | 3067 |
| $CH_3$ deformation | 1388.4 | 1440 |
| $CH_3$ deformation | 1468.1 | 1502 |
| C-C stretching | 994.8 | 1057 |
| $CH_3$ rocking | 1195.3 | 1202 |

However, both the parallel and perpendicular Raman spectra simulated by the BPM has produced abnormal signal at about 1202 cm$^{-1}$, inconsistent with the spectra simulated from PACFs by DFT and also consistent with the fact that the experimental intensity for the CH rocking peak is extremely weak[14,43]. This is due to the large errors in predicting PACF$_{aniso}$ by the BPM.

The most significant discrepancy between the Raman spectra simulated from PACFs by the AlphaML and DFT is the relative intensities for the CH$_3$ deformation at about 1440 and 1502 cm$^{-1}$, as shown in **Figure 4**. We owed the discrepancy to the limited ability of the AlphaML in predicting polarizability of structures that are far away from structures in the training data set. Although the AlphaML performs much better than the BPM in predicting polarizability anisotropy, as shown in **Figure 2c**, insufficient samplings[44] may still hinder the high-accuracy simulations of Raman spectra. An attempt of involving 40 structures from MD simulations into the training data set will produce a right prediction of the relative intensities for the CH$_3$ deformation at about 1440 and 1502 cm$^{-1}$, as shown in **Figure S3**. Moreover, **Figure S3** also demonstrates an excellent consistency between Raman spectra simulated from PACFs by the new trained AlphaML and DFT for 2-methylbutane.

**3.5 Discussions on predicting polarizabilities of larger molecules**

The additivity assumption for the molecular polarizability endows the AlphaML model and the BPM the ability with the ability of predicting polarizabilities for larger molecules. A benchmark testing was performed for alkanes containing no more than 11 carbons to evaluate the extrapolation ability of the AlphaML model and the BPM. We used the AlphaML model trained with 8 alkanes at equilibrium and additional 95 structures, with no non-equilibrium structures involved.

For alkanes containing no more than 5 carbons, which is already included in the training data sets, both the AlphaML and the BPM can predict $\bar{\alpha}$ and $\Delta\alpha$ as accurate as the DFT method, as shown in **Figure 5**. When it comes to alkanes containing more than 5 carbons, which is beyond the training data sets, large deviations happen for both the AlphaML model and the BPM. Especially, **Figure 5b** manifests that the points of $\Delta\alpha$ predicted by the BPM distributed irregularly and were far away from the y=x line, indicating that the BPM trained with small alkanes has very poor extrapolation ability. On the contrary, the points of $\Delta\alpha$ predicted by the AlphaML model deviated from the y=x line gradually. At the meantime, the deviations produced by the AlphaML model are much lower than those produced by the BPM.

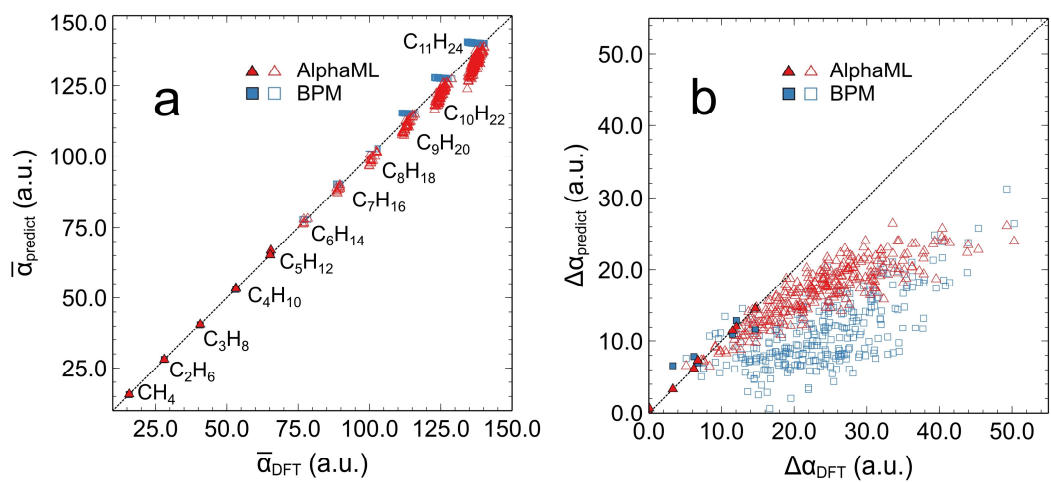

**Figure 5.** (a) Correlations between $\bar{\alpha}$ of alkanes containing no more than 11 carbons calculated by the DFT method and predicted by the AlphaML model and the BPM. (b) Correlations between $\Delta\alpha$ calculated by the DFT method and predicted by the AlphaML model and the BPM. Solid makers stand for alkanes containing no more than 5 carbons, while hollow markers stand for alkanes containing more than 5 carbons.

The lack of sampling certain local environment is responsible for the large deviations of $\Delta\alpha$ produced by the AlphaML model for alkanes containing more than 5 carbons. As shown in **Figure 6**, number of neighbors of the central carbon in n-$C_5H_{12}$ and n-$C_{11}H_{24}$ within the environment cutoff of 5 Å are 16 and 24, respectively. Therefore, the AlphaML model trained with alkanes containing no more than 5 carbons must be unfamiliar with the local environment of n-$C_{11}H_{24}$ and thus gives unsatisfied predictions of polarizabilities. **Figure 6** also suggests that a maximum of 9 carbons will be included within environment cutoff of 5 Å for n-alkanes. We inferred that the inclusion of structures containing 9 carbons in the training data set may be beneficial to the transferability of the AlphaML model to larger molecules such as n-$C_{11}H_{24}$. The consistency between $\bar{\alpha}$ and $\Delta\alpha$ calculated by DFT and predicted by the retrained AlphaML model in **Figure 7** clearly manifests that the complement of certain local environment with more neighbors will greatly improve the accuracy of predicting $\bar{\alpha}$ and $\Delta\alpha$, and thus endow the AlphaML with rational extrapolation ability.

Having shown the powerful ability of learning and predicting polarizability of the AlphaML model, future work will focus on the accurate prediction of polarizability of large systems such as polymers. This work clearly demonstrates the dependence of the accuracy of the AlphaML model on the diversity of the training data sets. Therefore, to guarantee the successful transference to polymers, sampling those local structures that suffused with atoms is indispensable[12].

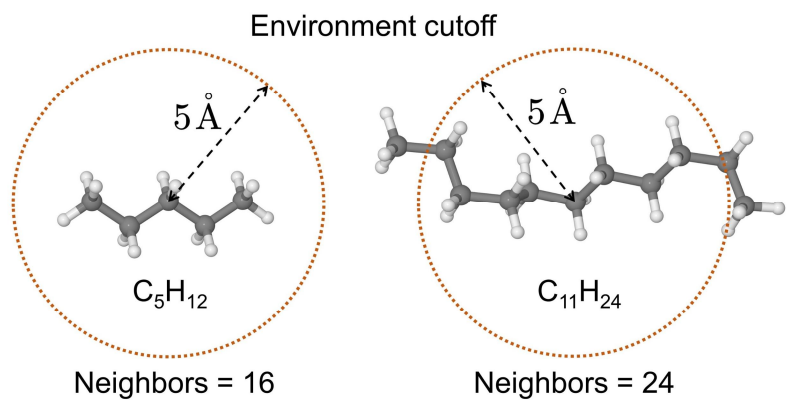

**Figure 6.** Number of neighbors of the central carbon in n-$C_5H_{12}$ and n-$C_{11}H_{24}$ within the environment cutoff of 5 Å.

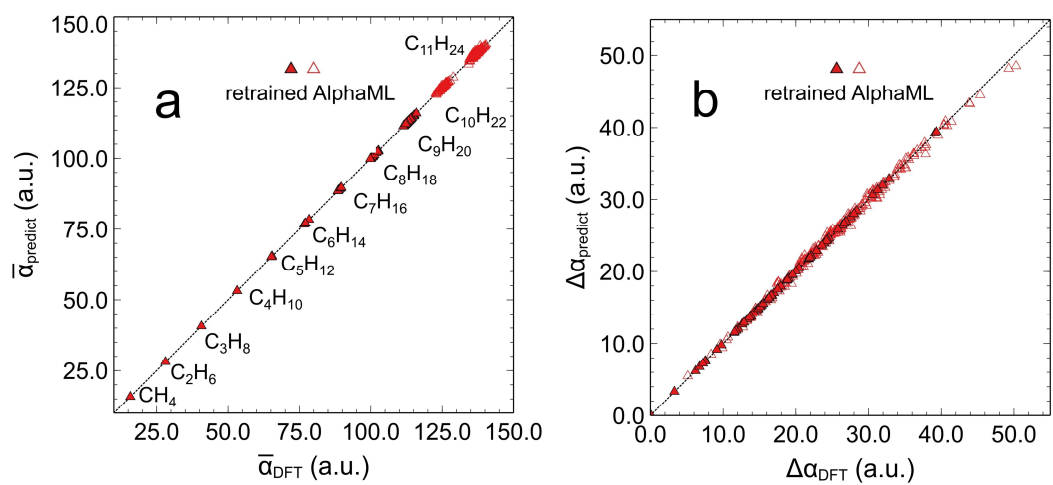

**Figure 7.** (a) Correlations between $\bar{\alpha}$ of alkanes containing no more than 11 carbons calculated by the DFT method and predicted by the retrained AlphaML model. (b) Correlations between $\Delta\alpha$ calculated by the DFT method and predicted by the retrained AlphaML model. Solid makers stand for alkanes containing no more than 9 carbons, while hollow markers stand for alkanes containing more than 9 carbons.

**4. Conclusion**

In this work, we constructed and compared the zero-order BPM and the ML-based AlphaML model for the prediction of polarizability and simulation of Raman spectra of alkanes. First, the BPM and the AlphaML were trained with polarizability of 8 alkanes at equilibrium together with the derivatives of polarizability associated with a bond stretch, respectively. The accuracy of these two models in the prediction of polarizability for molecules that are far away from equilibrium were compared. Then the time autocorrelation function of polarizabilities for $C_2H_6$ was calculated by these two models and the corresponding Raman spectra were simulated and compared with the Raman spectra by DFT. Finally, the extrapolation ability of these two models in predicting polarizability of alkanes larger than those in the training data sets were compared, and discussions were made for the AlphaML on the transference to large systems such as polymers.

We found that the AlphaML has appreciable advantages over the BPM in learning the polarizability and the derivative of polarizability of alkanes using the same training data set. Both the BPM and AlphaML can appropriately predict the isotropic polarizability for structures that are configurational different from those used in the training data sets. However, the BPM has inherent disadvantages in predicting polarizability anisotropy due to many factors including large uncertainties of estimating bond anisotropy, omitting of off-diagonal parameters in the expression of bond polarizability tensors. As a result, the BPM has large errors in the simulation of anisotropic Raman scattering. Finally, we demonstrated that both the BPM and AlphaML suffer from transference to alkanes larger than those used in the training data sets, but the problem for the AlphaML can be circumvented by enhancing samplings properly.

## AUTHOR INFORMATION

Corresponding Author

*E-mail: tamas@zju.edu.cn; yihezj@zju.edu.cn

## NOTES

The authors declare that there is no conflict of interest.

## ACKNOWLEDGEMENT

This work is supported by the National Key Research and Development Program of China (grant number 2022YFE0106100), and the National Natural Science Foundation of China (grant number 22178299). Nan Xu would like to thank the financial support provided by the Startup Funds of the Institute of Zhejiang University-Quzhou.## REFERENCES

(1) Wang, J.; Xie, X. Q.; Hou, T.; Xu, X. Fast Approaches for Molecular Polarizability Calculations. *The Journal of Physical Chemistry A* **2007**, *111* (20), 4443–4448. https://doi.org/10.1021/jp068423w.
(2) Mei, Y.; Yang, N.; Yang, W. Describing Polymer Polarizability with Localized Orbital Scaling Correction in Density Functional Theory. *The Journal of Chemical Physics* **2021**, *154* (5), 054302. https://doi.org/10.1063/5.0035883.
(3) Thomas, M.; Brehm, M.; Fligg, R.; Vöhringer, P.; Kirchner, B. Computing Vibrational Spectra from Ab Initio Molecular Dynamics. *Physical Chemistry Chemical Physics* **2013**, *15* (18), 6608. https://doi.org/10.1039/c3cp44302g.
(4) Berens, P. H.; White, S. R.; Wilson, K. R. Molecular Dynamics and Spectra. II. Diatomic Raman. *The Journal of Chemical Physics* **1981**, *75* (2), 515–529.
(5) Inakollu, V. S. S.; Geerke, D. P.; Rowley, C. N.; Yu, H. Polarisable Force Fields:24

# Supplementary information for: "A study of simulating Raman spectra for alkanes with a machine learning-based polarizability model"

*Mandi Fang,[a,b] Shi Tang,[a] Zheyong Fan,[c] Yao Shi,[b] Nan Xu,[a,b,*] Yi He,[a,b,d,*]*

a Institute of Zhejiang University-Quzhou, Quzhou 324000, China

b College of Chemical and Biological Engineering, Zhejiang University, Hangzhou 310027, China

c College of Physical Science and Technology, Bohai University, Jinzhou 121013, China

d Department of Chemical Engineering, University of Washington, Seattle, WA 98195, USA

# Contents

# 1 Principle of the bond polarizability model

This section presents the bond polarizability model (BPM) formalism briefly; detailed descriptions can be found elsewhere.[1] In the BPM, molecular polarizability tensor ($\boldsymbol{\alpha}$) is written as the sum of polarizabilities of all bonds

$$\boldsymbol{\alpha} = \sum_i \boldsymbol{\alpha}_i \tag{1}$$

where $\boldsymbol{\alpha}_i$ denotes the polarizability of bond $i$ in the fixed Cartesian axes. Supposed that $\mathbf{a}_i$ is the polarizability of bond $i$ in its principal axes, $\boldsymbol{\alpha}_i$ can be rewritten as

$$\boldsymbol{\alpha}_i = \mathbf{R}_i \mathbf{a}_i \mathbf{R}_i^{-1} \tag{2}$$

Here $\mathbf{R}_i$ is the rotation matrix between the bond's principal axes and the Cartesian axes. The bond's principal axes were constructed as follows: first, the longitudinal axis $L$ is directed along the vector connecting the bonded two atoms; Second, the two transversal axes $T_1$ and $T_2$ are built orthogonal to $L$ and to each other. $\mathbf{a}_i$ in the bond's principal axes reads

$$\mathbf{a}_i = \begin{pmatrix} a_L(i) & 0 & 0 \\ 0 & a_{T_1}(i) & 0 \\ 0 & 0 & a_{T_2}(i) \end{pmatrix} \tag{3}$$

where $a_L$, $a_{T_1}$, $a_{T_2}$ are the polarizabilities in the directions of the three principal axes.

In this work, we employed the zero-order BPM, which is commonly used in the predictions of polarizabilities.[1,2] The diagonal component of $\mathbf{a}_i$ ($a_s$, $s=L,T_1,T_2$) is expanded in a Taylor series respect to the bond length $r_i$ and truncated after the second term

$$a_s(i) = a_s^0(i) + a_s'(i) \cdot (r_i - r_i^0) + \cdots \tag{4}$$

Here the superscript 0 denotes the equilibrium state and the superscript ′ denotes the derivative of polarizability. In addition, a cylindrical bond model was used in the zero-order BPM, which assumes that $a_{T_1}$ and $a_{T_2}$ are identical.[1] Hence, we shall use $a_T$ to replace $a_{T_1}$ and $a_{T_2}$ in the following sections.

All bonds of the same type were described by the same set of parameters. The parameters $a_L^0$ and $a_T^0$ are known as the equilibrium parameters, which mainly account for the variation of polarizability upon a change of the bond's orientation.[1] We used the polarizabilities of 8 alkanes at equilibrium as the training data set to determine the four equilibrium parameters $a_{L(\text{C}-\text{C})}^0$, $a_{T(\text{C}-\text{C})}^0$, $a_{L(\text{C}-\text{H})}^0$, $a_{T(\text{C}-\text{H})}^0$ and the equilibrium bond length parameters $r_{(\text{C}-\text{C})}^0$, $r_{(\text{C}-\text{H})}^0$.

The differences between polarizabilities of molecules at equilibrium and molecules with a stretched C-C or C-H bond were then used to determine the derivative parameters $a_{L(\text{C}-\text{C})}'$, $a_{T(\text{C}-\text{C})}'$, $a_{L(\text{C}-\text{H})}'$ and $a_{T(\text{C}-\text{H})}'$. The fittings of $a_L^0$, $a_T^0$, $a_L'$ and $a_T'$ were performed with the use of the singular value decomposition method (SVD).[1]

## 2  Training of the bond polarizability model

The equilibrium parameters $a_{L(\text{C}-\text{C})}^0$, $a_{T(\text{C}-\text{C})}^0$, $a_{L(\text{C}-\text{H})}^0$ and $a_{T(\text{C}-\text{H})}^0$ in the BPM were fitted to be -49.571, 25.788, -27.361 and 14.556 Å$^3$, respectively. However, the negative values of $a_{L(\text{C}-\text{C})}^0$ and $a_{L(\text{C}-\text{H})}^0$ are not reasonable. The C-C and C-H bonds should increase the dielectric response in terms of the enhancement of the applied electric field along the bond's direction, which indicates that $a_{L(\text{C}-\text{C})}^0$ and $a_{L(\text{C}-\text{H})}^0$ should have positive values.[1,2] Moreover, the calculated bond anisotropy values of C-C and C-H bonds, defined as $\gamma^0 = a_L^0 - a_T^0$, are inconsistent with experimental results.[1,3,4] Zerbi and coworkers[4] reported that the C-H bond anisotropy $\gamma_{(C-H)}^0$ of methane was estimated to be 0.305 Å$^3$. Montero and coworkers[3] reported that $\gamma_{(C-C)}^0$ of ethane was about 1.28 Å$^3$ according to the experimental Raman spectra. The failure of the BPM in predicting the sign of polarizability along the bond's direction and the bond anisotropy may be due to the lack of intrinsic relations between $a_L^0$ and $a_T^0$ in the BPM model.

Imposing a restrictive condition for the C-H bond anisotropy in the BPM will remedy these shortcomings.[1] An addition equation defining the C-H bond anisotropy $\gamma_{(C-H)}^0 = 0.305$ Å$^3$ for the CH$_4$ molecule was introduced in the fitting of the equilibrium parameters. The new fit values of $a_{L(C-C)}^0$, $a_{T(C-C)}^0$, $a_{L(C-H)}^0$ and $a_{T(C-H)}^0$ are 1.678, 0.174, 0.777 and 0.484 Å$^3$, very close to the corresponding values of 1.677, 0.127, 0.779, 0.489 Å$^3$ in the work of Bougeard and coworkers.[1] The slight divergences may be due to the difference of the basis sets and QM packages. The bond anisotropy $\gamma_{(C-C)}^0$ and $\gamma_{(C-H)}^0$ were calculated to be 1.504 and 0.293 Å$^3$, close to the experimental $\gamma_{(C-C)}^0$ of 1.28 Å$^3$ for ethane and $\gamma_{(C-H)}^0$ of 0.305 Å$^3$ for methane.[3,4] The consistency between the polarizability tensors predicted by the BPM and calculated by the DFT method shown in **Figure S1a** confirms that the trained BPM can predict the polarizability of alkanes at equilibrium with an accuracy close to the DFT method. The R$^2$ for the diagonal components is 0.997; for the off-diagonal components, 0.803.

The best fit values of the derivative parameters $a'_{L(C-C)}$, $a'_{T(C-C)}$, $a'_{L(C-H)}$ and $a'_{T(C-H)}$ are 2.881, 0.274, 2.628 and 0.393 $\text{Å}^2$, in accord with the corresponding values of 2.863, 0.243, 2.743, 0.353 $\text{Å}^2$ in the work of Bougeard and coworkers.[1] In addition, the bond length parameters $r^0_{(C-C)}$ and $r^0_{(C-H)}$ were calculated to be 1.533 and 1.097 Å. **Figure S1b** demonstrates that the derivatives of polarizability associated with a bond stretch predicted by the BPM are in line with those calculated by the DFT method. The $R^2$ for the diagonal components of the derivatives of polarizability associated with a C-C bond stretch is 0.895; for the off-diagonal components, 0.879, while the $R^2$ for the diagonal components of the derivatives of polarizability associated with a C-H bond stretch is 0.842; for the off-diagonal components, 0.909.

## 3   Discussion on the large prediction errors of the bond polarizability model

Two reasons may contribute to the high RMSEs of polarizability anisotropy ($\Delta\alpha$) by the BPM. First, using the same set of parameters for all C-C bonds is not reasonable. The anisotropy of C-C bonds is found to be dependent on the local chemical environment, ranging from 0.6 to 1.4 $\text{Å}^3$.[3,5] The situation is also the same for C-H bonds. Second, the BPM has large errors in predicting the off-diagonal components of polarizability tensors. This may be originated from the omitting of off-diagonal parameters in the expression of bond polarizability tensors[6] and the omitting of second-order and higher-order terms in the Taylor series of parameters respect to the bond length.[1,6] However, an attempt to involve more parameters may cause too large statistical uncertainties.[6]

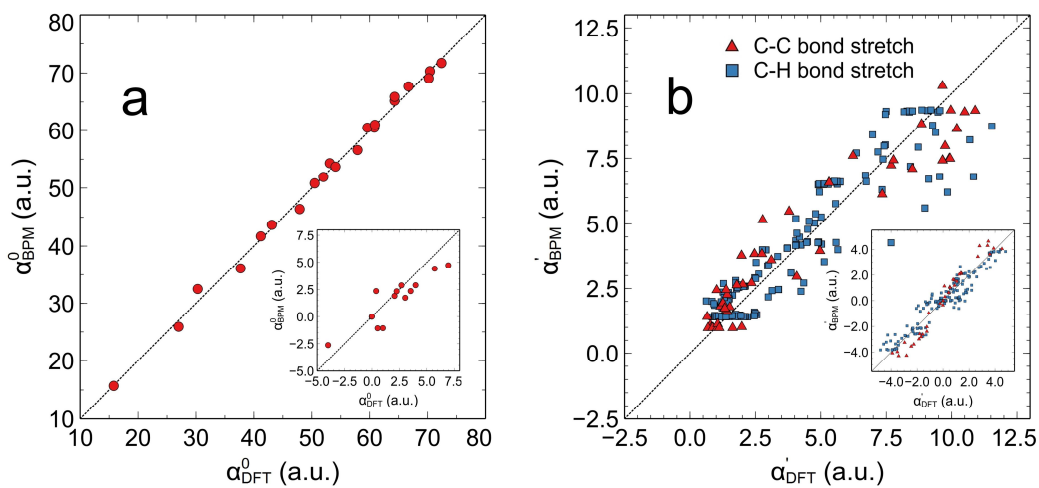

**Figure S1.** (a) Correlations between diagonal components of polarizabilities of alkanes at equilibrium calculated by the DFT method and predicted by the BPM. Correlations for off-diagonal components are shown in the inset. (b) Correlations between diagonal components of the derivatives of polarizability associated with a bond stretch calculated by the DFT method and predicted by the BPM. Correlations for off-diagonal components are shown in the inset.

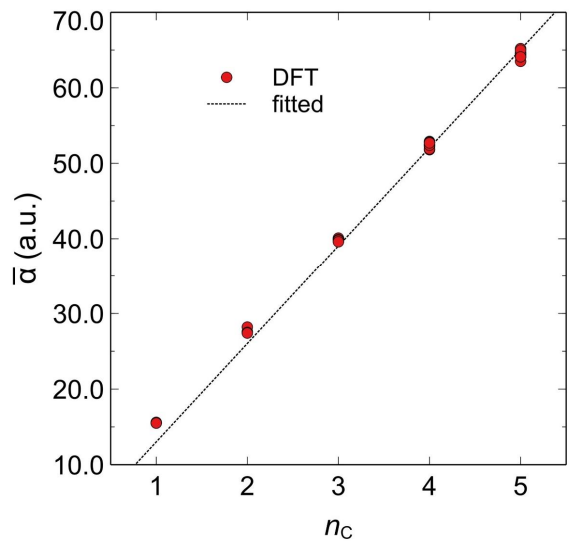

**Figure S2.** Correlations between isotropic polarizability ($\bar{\alpha}$) of non-equilibrium structures in the testing data set calculated by the DFT method and number of carbons ($n_\mathrm{C}$). The dashed line shows a fitted function of $y = 13.014x$.

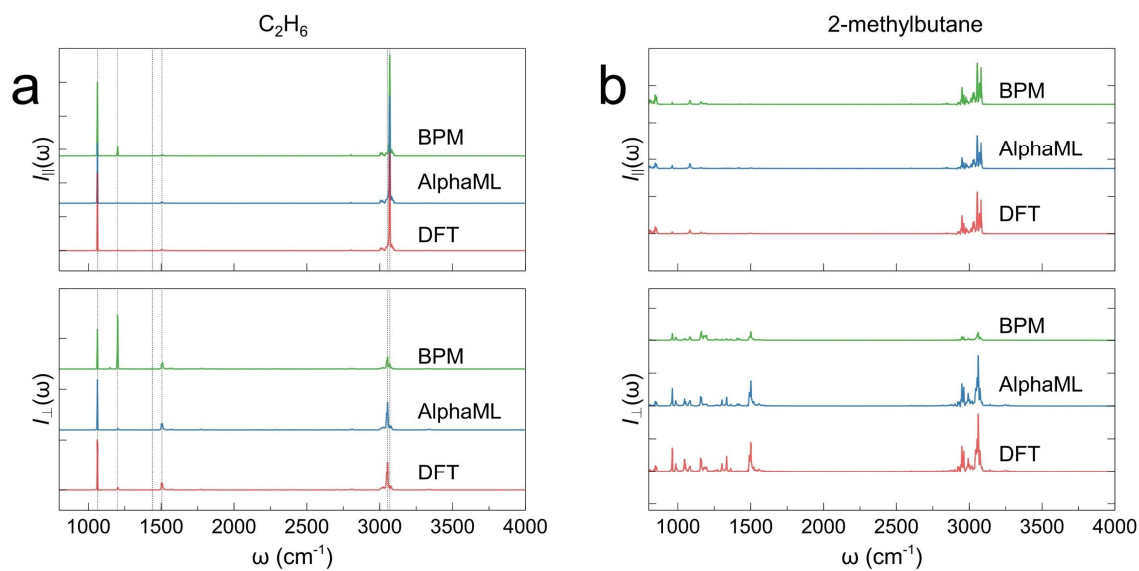

**Figure S3.** Comparisons between parallel and perpendicular Raman spectra of (a) $C_2H_6$ and (b) 2-methylbutane simulated from PACFs by the DFT method, the AlphaML and the BPM. The latter two Raman spectra are shifted upward. The training data set for AlphaML includes polarizabilities of 8 alkanes at equilibrium, derivatives of polarizability associated with a bond stretch and polarizabilities of 40 structures from MD simulations.